\documentclass{WileyMSP-template}
\usepackage{color}

\begin{document}

\pagestyle{fancy}
\rhead{\includegraphics[width=2.5cm]{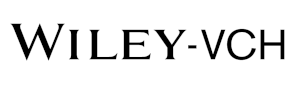}}

\title{Digital-Analog Quantum Simulation and Computing: A Perspective on Past and Future Developments}

\maketitle

% Author: Please give full first and last names for authors and include * after the name of all corresponding authors

\author{Lucas Lamata}*

% Dedication

%\dedication{Dedicated to the founders of Quantum Mechanics}

% Affiliations: Please provide adacemic titles (Prof. or Dr.) for all authors where applicable, and include an institutional email address for all corresponding authors
\begin{affiliations}
L. Lamata\\
Departamento de F\'isica At\'omica, Molecular y Nuclear, Facultad de F\'isica, Universidad de Sevilla, Apartado 1065, E-41080 Sevilla, Spain\\
Email Address: llamata@us.es

\end{affiliations}

% Keywords: Please provide a minimum of three and a maximum of seven keywords, separated by commas


% Abstract should be written in the present tense and impersonal style (i.e., avoid we), and be at most 200 words long
\begin{abstract}

Quantum simulation and computing traditionally has been based on two main paradigms, namely, digital and analog. In the digital paradigm, usually single and two-qubit gates (where qubit is an acronym for quantum bit) are employed as building blocks for scalable, universal quantum computing, although errors add up fast and error correction will be ultimately needed for scaling up. In the analog paradigm, large analog blocks are normally employed for a unitary dynamics that carries out the computation, enabling quantum operations on many qubits with reduced errors, but with the drawback of a limited choice of evolutions and lack of universality. In the past decade, a new paradigm has emerged, showing interesting possibilities for quantum simulation and computing in the near and mid term. This is the paradigm of digital-analog quantum technologies, which proposes to combine the best of both paradigms: large analog blocks, provided by native interactions of the employed quantum platform, enabling scalability, combined with digital gates, allowing for more versatility and, ultimately, universality. In this Perspective, I give an overview of the evolution of the field along the past decade, and an outlook for its future possibilities.  
\end{abstract}

% Text: Please use section headings and subheadings as specified below. For communications, all section headings apart from Experimental Section should be removed
% Please make the first reference to a display item bold: \textbf{Figure 1}
% Do not abbreviate Figure, Equation, etc.; display items are always singular, i.e., Figure 1 and 2.
% Equations are always singular, i.e., Equation 1 and 2, and should be inserted using the {equation} environment, not as graphics
% Please do not use footnotes in the text, additional information can be added to the Reference list.

\section{Digital-Analog Quantum Technologies}

Since the late 20th century, the field of quantum computing emerged as a promising new computing paradigm for solving problems unattainable by the standard computing architectures [1]. This area considered genuine quantum platforms such as trapped ions, superconducting circuits, and cold atoms, among others, for carrying out exponentially faster computations for, among others, breaking cryptographic encodings, and designing novel materials. Theoretically, this was shown to be feasible, as far as one could build full-fledged scalable quantum computers. In practice, however, quantum systems are fragile, due to the phenomenon of quantum decoherence, and it is utterly difficult to control quantum platforms with the necessary degree to build a fully functional quantum computer with several hundreds of qubits. Theoreticians aimed at this challenge via the design of quantum error correction protocols, already in the 90's [1]. But the truth is that the necessary overhead in the number of physical qubits to build a useful set of logical qubits, has made this prospective error-corrected quantum architecture impractical so far. Therefore, the dream of a fully functional, scalable quantum computer, seems out of reach as of today.

\vspace{1cm} 

About 15 years ago, theoreticians started working on a novel quantum computing paradigm [2,3], see Fig. 1. After realizing that digital protocols for quantum simulation and computation would take many years to be really useful, new pragmatic approaches were pursued, that crystallized in the digital-analog quantum technologies paradigm. In this paradigm, first addressing the field of quantum simulations, the aim was to employ the native interaction in a given platform, for carrying out many-qubit gates with an always-on driving, and combine it with single-qubit gates, perhaps adding also some two-qubit gates as needed. The motivation for this approach was to carry out large entangling blocks on as many qubits as feasible, in order to have a scalable interaction acting on tens or even hundreds of qubits, combined with single and perhaps a few two-qubit gates, to allow for a wider variety of dynamics and ultimately a universal, scalable quantum paradigm. The initial hope was to be able to address a useful set of problems with current or near future quantum devices, with no need to resort to quantum error correction paradigms. For early reviews in the field, in the context of quantum platforms of trapped ions and superconducting circuits, see Refs. [2,3].

\vspace{1cm} 

In more recent years, a variant of the previous schemes addressing universal quantum computing was put forward [4], and the combination with the quantum machine learning area, with a feasible implementation via digital-analog quantum techniques, was also analyzed [5].  It should also be highlighted that the digital-analog quantum simulation and computing paradigm includes as particular cases the analog and digital ones, such that it is automatically a universal paradigm for quantum computing resorting to the universality proofs in the purely digital case of quantum computing. Namely, by having a universal set of single and two-qubit gates [1], adding analog blocks for speeding up and simplifying the computation does not reduce the universal character of the quantum computer, it just makes the experimental implementation more efficient given the current available technology [2,3].

\begin{figure}
\center
\includegraphics[width=0.5\linewidth]{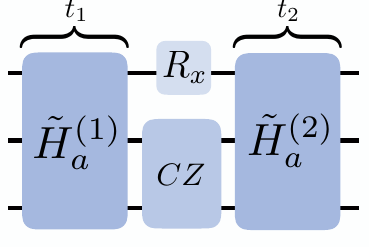}
\caption{Scheme of an instance of a basic digital-analog quantum computation. It consists of digital gates, here represented as a single-qubit $R_x$ rotation and a two-qubit Controlled-Z $CZ$ gate, and analog blocks given by time evolutions of the quantum simulator with native interactions, in this particular example provided by the Hamiltonians $\tilde{H}_a^{(1)}$ evolved a time $t_1$, and $\tilde{H}_a^{(2)}$ evolved a time $t_2$. Adapted from Ref. [3].}
\end{figure}

\vspace{1cm} 

In this Perspective, I give an overview of the past developments in the digital-analog quantum simulation and computing field, and give also an outlook on the future possibilities for this promising field. I start by reviewing the area of digital-analog quantum simulations (Sect. 2), to then describe the digital-analog quantum computing and machine learning topics (Sect. 3). In Sect. 4, I explore novel approaches followed by theoreticians in the digital-analog quantum realm, that aim at proving quantum advantage, and error correction techniques, inside this growing quantum paradigm, from a more mathematical approach. In Sect. 5, I give a non-exhaustive overview of the many significant quantum experiments carried out in the past five years, with the digital-analog quantum paradigm or similar ones, in the quantum platforms of trapped ions, superconducting circuits, and cold atoms (e.g., Rydberg atoms), reaching many dozens or hundreds of qubits, and starting to experimentally address important novel problems of, e.g., quantum materials. Finally, I give an outlook for the field in Sect. 6, with my personal take on what the future may hold in this area.

\section{Digital-Analog Quantum Simulations}

Quantum simulations were proposed among others by Richard Feynman, already in the 80's, and since then have progressed to a significant extent, constituting one of the most developed and fruitful fields inside quantum technologies [6]. A quantum simulation is a controlled reproduction of the dynamics and/or static properties of a given quantum system under study, by another, controllable, quantum system that one may use as a quantum simulator of the former. The idea, originally by Feynman, is that a quantum system is often more useful to emulate another quantum system, given that the Hilbert space dimension of large quantum systems grows exponentially with the number of quantum particles, atoms, etc., and this makes the computation unfeasible with standard computers for around 50 to 60 qubits, at least in cases with no obvious symmetry.

 \vspace{1cm}

Traditionally, two paradigms for quantum simulations have been pursued, namely, digital quantum simulations, and analog quantum simulations [6]. As mentioned above, digital quantum simulations aim at having a universal quantum platform, in this case in order to be able to simulate any quantum system of similar dimensionality. In standard digital approaches, one would decompose a given quantum dynamics via digital steps, with a Lie-Trotter-Suzuki decomposition, which could in principle reproduce the quantum evolution of any quantum system. The drawback with current quantum platforms is that with a purely digital approach it is really hard to go beyond, say, 10 qubits nowadays, both in the simulated quantum system and in the simulator, as the errors in the digital gates add up fast rendering the simulation unfeasible when pushing it further.

\vspace{1cm}

Analog quantum simulations have been another avenue pursued in this field, in which the dynamics is typically given by a controllable Hamiltonian (i.e., native interaction) of the quantum platform employed as simulator, which aims at matching as far as possible the simulated quantum system [6]. This approach has been successful in a series of quantum platforms, such as trapped ions and cold atoms, but it has as drawback that it can simulate only a reduced number of quantum systems, typically the ones that are similar to the quantum simulator platform, and this means that often the problems that can be addressed are not highly interesting scientifically and/or technologically.

\vspace{1cm}

Since about 2011-2012, theoreticians started working in another quantum paradigm, merging both the digital and analog quantum simulation paradigms, in order to carry out more efficient, versatile, and scalable quantum simulations. The aim, as mentioned above, was to benefit from the scalable properties of native analog blocks on many qubits, and the versatility and universality of digital, few qubit gates applied on the analog dynamics, or interspersed with it. This research line started with proposals for quantum simulations in trapped ions, and stemmed from an original theory proposal [7] in 2007 for the quantum simulation of the Dirac equation with a single trapped ion, which, being itself an analog quantum simulator proposal, already contained the seed for the digital-analog approach, as it involved a two-level quantum system emulating the relativistic electron spinor, and a bosonic mode to emulate the momentum of the Dirac electron, namely, a discrete and a continuous degree of freedom.
Years later, proposals for quantum simulations with trapped ions of quantum field theories, fermionic lattice models, and fermionic-bosonic quantum systems, involving, progressively, digital decompositions and large analog blocks with the vibronic modes in the native trapped ion quantum platform, were put forward. At this point, the motivation was largely to propose how to carry out efficient quantum simulations of fermionic systems with digital and digital-analog quantum simulators in trapped ions, which was already an unsolved problem in more than one spatial dimension. This was the period roughly between 2011 and 2014, which was summarized in the review paper in Ref. [2].

\vspace{1cm}

A subsequent period between 2014 and 2018 involved further proposals for quantum simulations with the platform of superconducting circuits, such as the digital-analog quantum simulation of the quantum Rabi model, among others, as well as similar proposals of quantum simulations of fermionic and bosonic degrees of freedom in this other quantum platform, adapted from the ones of trapped ions, for studying condensed matter, high energy and quantum chemistry systems. Collaborations with prominent experimental groups in this period evidenced that a purely digital approach would not produce useful and practical results in quantum simulations unless the technology was significantly enhanced. It was at this time where it became more apparent that the digital-analog quantum paradigm would be one promising way to produce useful results and learning new physics in quantum systems in the near and mid term. The review of Ref. [3] summarizes the research carried out along these lines, which consolidated this paradigm at the theory level.

\section{Digital-Analog Quantum Computing and Machine Learning}

After the digital-analog quantum simulator paradigm was basically established theoretically, a further analysis was carried out in order to more formally establish universal properties of the digital-analog quantum paradigm from the quantum computing (not only simulation) point of view. Here, the aim was to start with an analog native interaction, such as an Ising or XY spin model in a large quantum system and apply single-qubit gates on it in order to be able to produce harder quantum models in large quantum systems, such as, e.g., a Heisenberg model. This was already shown for quantum simulations with trapped ions in simple instances, but in this case it was more formally studied, and the universal character was given mathematical rigor, as shown in Ref. [4].

\vspace{1cm}

More recently, studies of the possible implementation of quantum machine learning protocols with the digital-analog quantum paradigm have been reviewed in another Perspective in Ref. [5].

\section{Towards Quantum Advantage with the Digital-Analog Quantum Paradigm}

In the past few months, extensive theoretical and mathematical analyses inside the digital-analog quantum paradigm have been published, which point to this field as a promising one both for achieving a quantum advantage with current quantum platforms, and to efficiently implement quantum error correction techniques in digital (i.e., qubits)-analog (i.e., quantum oscillators) systems.

\vspace{1cm}

In Ref. [8], a rigorous mathematical study of quantum advantage with a digital-analog quantum computing system is performed. Namely, a quantum system composed of a large analog block with a transverse field Ising interaction on many qubits, with single qubit gates implemented before and after the analog block, and subsequent Z basis quantum measurement, is shown to already enable a provable speedup with respect to classical paradigms. Specifically, the paper states that ``Quantum supremacy tests are possible on today's quantum annealers, as well as other devices capable of hybrid digital-analog quantum evolution''.

\vspace{1cm}

Moreover, Ref. [9] describes an extensive theoretical framework for a particular case of a digital-analog quantum scenario, which involves quantum bits (digital elements) and quantum oscillators (analog elements) as well as interactions among them. In this large analysis, with deep mathematical approach, issues such as error correction techniques applied in this kind of paradigm are addressed, as well as quantum speedup considerations. 

\vspace{1cm}

These examples are just further evidence that from a purely theoretical and computer science point view the digital-analog quantum simulation and computing paradigm seems promising and highly fruitful to develop. In the next section, I will just give further evidence of this fact, with a non-exhaustive summary of the plethora of world-record quantum experiments carried out in the past 5 years or so, in quantum platforms such as trapped ions, superconducting circuits, and cold (Rydberg) atoms.

\section{Implementations of the Digital-Analog Quantum Paradigm: Current Status}

In the past 5 years, a significant amount of quantum experiments in quantum platforms such as trapped ions, superconducting circuits, and cold atoms, have beaten all previous records of numbers of quantum bits and quantum gates performed, reaching often many dozens or even hundreds of quantum bits for useful quantum simulations and computations. This has been enabled mainly because of the combination of two key issues: Firstly, the technological advances in all these three platforms, in terms of qubit stability and coherence, and gate fidelities, have made these experiments a reality. Second, in most cases, these experiments have achieved the record-number qubits employed for useful computations via digital-analog quantum techniques, or some alternative version of this kind of paradigm. The interesting thing, in my view, of the digital-analog quantum paradigm, is that i) it works in all three platforms, and ii) no platform would achieve, in general, these record numbers of qubits and complexity of experiments, nowadays, without using the digital-analog quantum paradigm or a similar approach. This is also highlighting the relevance, at least in the short and mid term, of the digital-analog quantum paradigm.

\vspace{1cm}

In the following, and given the many experiments already carried out in the past few years in this context, I will describe a few examples in each platform, just to give a flavor of the usefulness of this approach, instead of exhaustively reviewing the whole experimental quantum literature on this topic. Being this a Perspective for Advanced Computing, with, therefore, a readership with interests in computing in general, and not specifically on quantum technologies, I prefer to be concise in this aspect, and refer to the search engines and citation lists of some of these experiments, which, in some cases, already go above several hundreds of citations in a few years.

\vspace{1cm}

I will divide this section on three subsections, for trapped ions, superconducting circuits, and cold atoms, being these three the quantum platforms with enough technological developments such as to enable digital-analog quantum implementations with at least many dozens of qubits.

\subsection{Trapped Ions}

The technology for large scale quantum simulations with trapped ions involving long chains of dozens of trapped ions, and digital as well as analog control, has mainly been developed in the Innsbruck group along many decades. In this group the quantum simulation of the Dirac equation proposed by us in 2007, Ref. [7], was carried out in 2010 [10], in an important analog quantum simulation experiment at the time. Pioneering experiments of digital quantum simulations involving a few trapped ions were carried out already in 2011 in this trapped ion lab [11], at the time where theoretical ideas mixing digital and analog protocols in trapped ions were starting to be proposed, partially motivated by this experiment, as summarized in Ref. [2]. In the past 5 years or so, trapped ions has become one of the quantum platforms where experiments involving dozens of qubits with digital and analog control, or similar techniques, have been carried out. In the Innsbruck group, one particular pioneering example is Ref. [12]. 

\vspace{1cm}

The group of Duke University has also achieved digital-analog quantum simulations with dozens of qubits in trapped ion experiments, using similar techniques as the Innsbruck group, mostly in quantum simulation of physical problems of relevance. One particular recent example is Ref. [13].

\vspace{1cm}

Besides the previous groups in Europe and the US, the Chinese group of Prof. Kihwan Kim in Tsinghua University has also produced interesting trapped ion experiments recently with the digital-analog quantum simulation paradigm, as in Ref. [14].

\subsection{Superconducting Circuits}

The quantum platform of superconducting circuits has recently achieved impressive quantum experiments using digital-analog quantum technologies, for example in the Google Quantum AI lab [15], as well as the D-Wave Quantum Computing Company [16], in the first case, employing qubits and global interactions, and in the second case, a quantum annealer as commonly used in D-Wave complemented with single-qubit gates. An earlier, pioneering experiment in the Chinese lab of Prof. Jian Wei Pan has also been published in 2023 [17].These experiments, focused mainly on quantum simulations, are already starting to produce useful knowledge in the realm of quantum materials, for example, often involving more than 50 qubits for useful quantum computations. The Google Quantum AI lab inherited the technology developed by Prof. John Martinis and coworkers in University of California Santa Barbara along the previous decades, which was recognized with the Nobel Prize in Physics 2025 for  Prof. Martinis.

\subsection{Cold Atoms}

Perhaps, among the three quantum platforms achieving impressive numbers of qubits and experiment complexity in the past few years, employing digital-analog quantum techniques or similar ones, cold atoms is the most surprising one to me, because for many decades it was the most scalable quantum simulator platform, but just in the analog context, while, on the contrary, digital protocols were unfeasible in this platform for many years because of the low two-qubit gate fidelities.

\vspace{1cm}

However, in the past 5 years, or so, high-fidelity experiments with Rydberg states employing digital-analog quantum control and technologies, have achieved numbers of atoms that in some cases not only go above dozens, but even hundreds, making this quantum platform a highly significant one, nowadays, for carrying out scalable quantum simulation and computation for problems that may be useful to solve and at the same time unfeasible with standard classical computers.

\vspace{1cm}

Some examples of experiments and groups having currently this technology are the Harvard cold atom group, which has pioneered experiments with Rydberg atoms at this level of complexity, as evidenced in Refs. [18-20], and the related startup company QuEra, as well as the Pasqal startup in Paris [21,22].

\section{Outlook}

In this Perspective, I have made a personal account of the emergence and evolution of the digital-analog quantum paradigm. It is, therefore, not intended to be an exhaustive review. This kind of approach seems to work well in all three platforms which have enough technological developments as to quantum control a sufficiently large number of qubits to be able to prove that the approach works. Moreover, any quantum experiment aiming at carrying out a quantum simulation or computation with at least several dozen qubits, at the same time as addressing quantum calculations which are not somehow trivial, or previously known, has typically to employ the digital-analog quantum paradigm nowadays, or some similar approach.

\vspace{1cm}

Since the pioneering proposal for quantum computations with cold trapped ions of Cirac and Zoller in 1995 [23], much progress in the field has been produced, and it is rewarding that experiments are reaching a degree of control enabling to employ many tens or even hundreds of quantum bits for carrying out meaningfully quantum simulations and computations for interesting quantum problems. It is also inspiring that the digital-analog quantum technology paradigm is playing an important role in this possibility. As far as the technology still does not allow one for developing a full-fledged quantum error corrected computer, one may conjecture that the digital-analog quantum paradigm may still be a fruitful avenue for years to come.

% Acknowledgements
\medskip
\textbf{Acknowledgements} \par %delete if not applicable))

I acknowledge support from grants PID2022-136228NB-C21 and PID2022-136228NB-C22 funded by
\\ MCIN/AEI/10.13039/50110001103 and “ERDF A way
of making Europe”, and by the Ministry for Digital Transformation and of Civil Service of the Spanish Government
through the QUANTUM ENIA project call - Quantum
Spain project, and by the European Union through
the Recovery, Transformation and Resilience Plan -
NextGenerationEU within the framework of the “Digital
Spain 2026 Agenda”.

% References
\medskip

% Use the following code if you wish to generate your bibliography with BibTeX;
% replace the string "MSP-template" below with the name(s) of
% the BibTeX data base(s) you want to use.
% The resulting bibliography-output (the content of the .bbl file)
% must be pasted back into this file before submission.
% Please also include your BibTeX data base file(s) in your submission
% so that we can re-run BibTeX if necessary.
%
%\bibliographystyle{MSP}
%\bibliography{MSP-template}

\textbf{References}\\

1 Michael A. Nielsen and Isaac L. Chuang, Quantum Computation and Quantum Information, Cambridge University Press, Cambridge, UK (2000).\\

2 L. Lamata, A. Mezzacapo, J. Casanova, and E. Solano, Efficient quantum simulation of fermionic and bosonic models in trapped ions, EPJ Quantum Technology 1, 9 (2014).

3 L. Lamata, A. Parra-Rodriguez, M. Sanz, and E. Solano, Digital-Analog Quantum Simulations with Superconducting Circuits, Adv. Phys.: X 3 (1), 1457981 (2018).

4 A. Parra-Rodriguez, P. Lougovski, L. Lamata, E. Solano, and M. Sanz, Digital-analog quantum computation, Phys. Rev. A 101 (2), 022305 (2020).

5 L Lamata, Digital-Analog Quantum Machine Learning, Advanced Intelligent Discovery, 1 (1), 2400023 (2025).
	
6 Andrew J. Daley, Immanuel Bloch, Christian Kokail, Stuart Flannigan, Natalie Pearson, Matthias Troyer and Peter Zoller, Practical quantum advantage in quantum simulation, Nature 607, 667 (2022).

7 L. Lamata, J. Le\'on, T. Sch\"atz, and E. Solano, Dirac Equation and Quantum Relativistic Effects in a Single Trapped Ion, Phys. Rev. Lett. 98, 253005 (2007).

8 D. Lidar, Digital-Analog-Digital Quantum Supremacy, arXiv:2512.07127 (2025).

9 Y. Liu et al., Hybrid Oscillator-Qubit Quantum Processors: Instruction Set Architectures, Abstract Machine Models, and Applications, PRX Quantum 7, 010201 (2026).

10 R. Gerritsma et al., Quantum simulation of the Dirac equation, Nature 463, 68 (2010).

11 B. P. Lanyon et al., Universal digital quantum simulation with trapped ions, Science 334, 57 (2011).

12 M. K. Joshi et al., Exploring large-scale entanglement in quantum simulation, Nature 624, 539 (2023).

13  L. Feng et al., Continuous symmetry breaking in a trapped-ion spin chain, Nature 623, 713 (2023). 

14 Y. Lu et al., Implementing Arbitrary Ising Models with a Trapped-Ion Quantum Processor, Phys. Rev. Lett. 134, 050602 (2025).

15 T. I. Andersen et al., Thermalization and criticality on an analogue-digital quantum simulator, Nature 638, 79 (2025).

16 R. Deshpande et al., Analog-Digital Quantum Computing with Quantum Annealing Processors, \\ arXiv:2603.15534 (2026).

17 M. Gong et al., Quantum neuronal sensing of quantum many-body states on a 61-qubit programmable superconducting processor, Sci. Bull. 68, 906 (2023).

18 H. Bernien et al., Probing many-body dynamics on a 51-atom quantum simulator, Nature 551, 579 (2017).

19 T. Manovitz et al., Quantum coarsening and collective dynamics on a programmable simulator, Nature 638, 86 (2025).

20 A. A. Geim et al., Engineering quantum criticality and dynamics on an analog-digital simulator, \\ arXiv:2602.18555 (2026).

21 Pasqal, www.pasqal.com

22 A. Michel et al., Blueprint for a digital-analog variational quantum eigensolver using Rydberg atom arrays, Phys. Rev. A 107, 042602 (2023).

23 J. I. Cirac and P. Zoller, Quantum Computations with Cold Trapped Ions, Phys. Rev. Lett. 74, 4091 (1995).

% Please provide Biographies and photos for Essays, Feature Articles, Progress Reports, Reviews, and Perspectives for those authors who should be highlighted  
% These should be at most 100 words long
% For other article types this section can be removed
% Photographs should be 40mm broad and 50 mm high

% Table of contents entry should be 50 - 60 words long
% Image should be 55 mm broad and 50 mm high or 110 mm broad and 20 mm high

%\begin{figure}
%\textbf{Table of Contents}\\
%\medskip
 % \includegraphics{toc-image.png}
 % \medskip
 % \caption*{ToC Entry}
%\end{figure}

\end{document}